\documentclass[sort&compress]{aipproc}

\layoutstyle{6x9}

\begin{document}

\title{Gas accretion by planetary cores}


\author{Ben A. Ayliffe}{
  address={School of Physics, University of Exeter, Stocker Road, Exeter, EX4 4QL}
}

\author{Matthew R. Bate}{
}

\begin{abstract}
We present accretion rates obtained from three-dimensional self-gravitating radiation hydrodynamical models of giant planet growth. We investigate the dependence of accretion rates upon grain opacity and core/protoplanet mass. The accretion rates found for low mass cores are inline with the results of previous one-dimensional models that include radiative transfer.

\end{abstract}

\maketitle

\section{Introduction}

The core accretion model of planet growth has been devised to explain the formation of giant planets such as Jupiter and Saturn {\cite{PerCam1974, Mizuno1978}. Through the coagulation of solid grains, planetesimals are built which then, through a so far uncertain process, bind to form large km scale bodies. When such bodies collide they shatter, but more often than not are held together by their self-gravity \cite{Safronov1969}, allowing them to grow through collisions to reach many Earth masses. These large solid masses exert a considerable gravitational tug on the solar nebula in which they are immersed, causing them to slowly gather a gaseous atmosphere. If the solid cores are sufficiently massive (> 10~$\rm M_{Earth}$) they may trigger runaway gas accretion. This occurs when the accreted gaseous atmosphere obtains a mass similar to the solid core, causing it to collapse from its hydrostatic state. The collapse leads to a rapid in fall of material from the surrounding nebula to replace the condensing gas. It is believed that such a process allows the accretion of hundreds of Earth masses of gas.

Three-dimensional hydrodynamical calculations examining the interaction of a protoplanet with a disc have usually assumed a locally isothermal equation of state, and/or do not resolve the gas flow far inside the planet's Hill radius \citep{Brydenetal1999,Kley1999,LubSeiArt1999,Nelsonetal2000,Masset2002,batezeus,DAnHenKle2003,DAnBatLub2005,PaaMel2006,DAnLubBat2006}. Recent hydrodynamical calculations have also considered different equations of state and/or radiation transport \citep{DAnHenKle2003, MorTan2003, KlaKle2006, PaaMel2006, PaaMel2008}. Fouchet \& Mayer \cite{FouMay2008} include both radiative transfer and self-gravity, but once again the resolution of gas flow near to the planetary core is limited.
In this work we present results from three-dimensional self-gravitating radiative transfer hydrodynamical calculations that are able to resolve gas flow down to realistic core radii. We vary the core/protoplanet masses and the grain opacity of the nebula to examine the effects that these properties have upon the gas accretion rates of the growing planets.

\section{Model setup}

We model 10 - 333 $\rm M_{Earth}$ solid cores/protoplanets embedded in a circumstellar disc with a surface density of 75~$\rm g~cm^{-2}$ orbiting a 1$\rm M_{\odot}$ star at 5.2 AU. The initial surface density has a $\Sigma \propto r^{-1/2}$ profile, whilst the initial temperature profile of the disc is $T \propto r^{-1}$, yielding a constant disc scaleheight with radius of $H/r=0.05$. We vary the grain opacity of the disc from the interstellar grain opacity (IGO) down to 0.1\% of this value to examine the effects this has upon the disc cooling.

To achieve high resolution near a protoplanet at reasonable computational cost we model only a small section of the protoplanetary disc centred on the planetary core. Our section size is $5.2 \pm 0.78~{\rm AU}$, and $\phi = \pm 0.15$ radians; for details see \cite{AylBate2009}.

\section{Results \& Discussion}

The accretion rates for the different mass cores/protoplanets that we modelled under differing opacities are shown in the left panel of figure \ref{figure1}. The trend in accretion rate against core/protoplanet mass is as would be expected, increasing as the central bodies mass increases. There is a reduction in accretion rates for the highest mass protoplanets caused by the evacuation of a large gap in the encompassing circumstellar disc by these protoplanets. Such a gap reduces the material available in the feeding zone of the protoplanet. The accretion rates also show an increasing trend with reduced grain opacities for the low mass cores. Reduced grain opacities allow the infalling material to cool more readily via radiation, and so the thermal pressure supporting it against capture is reduced. The only exception to this is at the lowest grain opacity of 0.1\% IGO for the lowest mass core where the radiation and gas fields interact so little that the two fields decouple, preventing effective gas cooling via radiation. For the high mass protoplanets the grain opacity has very little impact on the accretion rates as the process is gravitationally dominated.

The right hand panel of figure \ref{figure1} compares the accretion on to a 10$\rm M_{Earth}$ solid core in our models with the accretion onto 5 and 15 $\rm M_{Earth}$ cores in the one-dimensional models of Papaloizou \& Nelson \cite{papa2005}. It can be seen that the results show good agreement, suggesting that for at least the phase of accretion preceding runaway growth, the transition from one-dimensional models to three-dimensional models does not have a significant impact on the accretion rates.

Figure \ref{figure2} illustrates the differences between radiation hydrodynamical and isothermal calculations. Features in the vicinity of the planet that are well established under isothermal conditions, such as spiral shocks and circumplanetary discs, are weakened in the radiative transfer calculations. The thermal pressure acts to smear out these features. It can also be seen in figure \ref{figure2} that reducing the grain opacity leads to a model that is more similar to the isothermal case. The increased cooling seen at lower opacities reduces the thermal pressure, thus mitigating its impact on the shocks and circumplanetary discs.

\begin{figure}
  \begin{minipage}[b]{0.5\linewidth} 
    \centering
    \includegraphics[height=6.5cm]{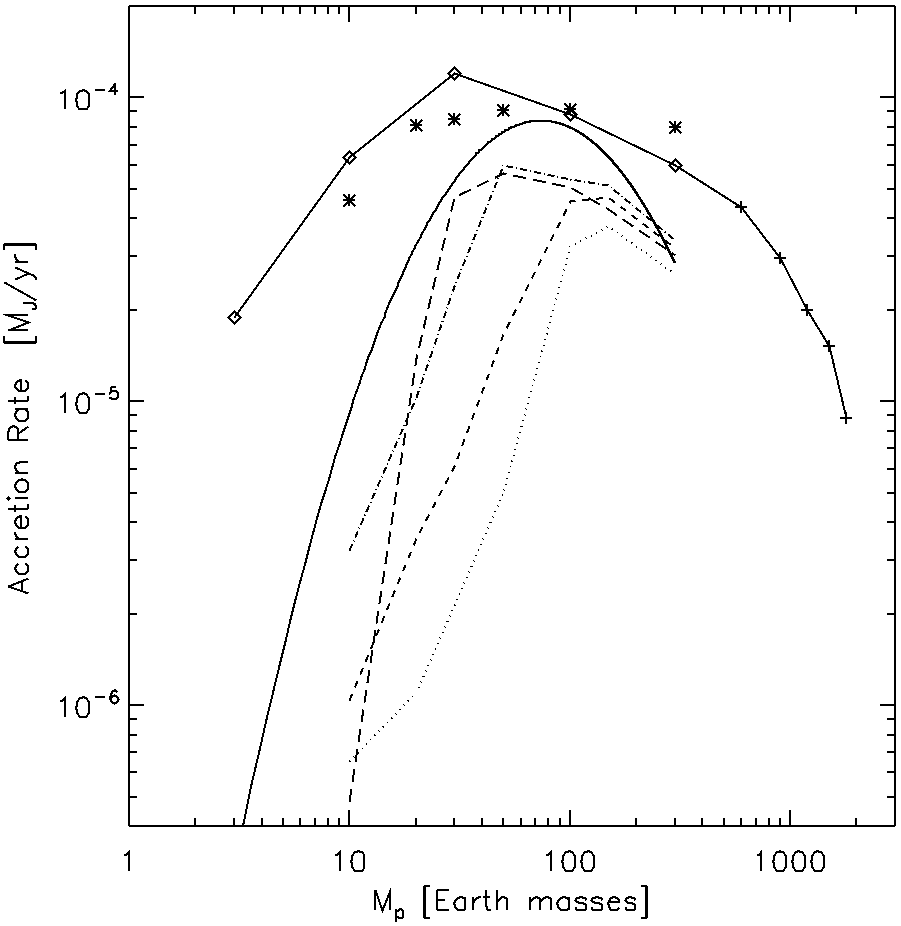}
  \end{minipage}
  \hspace{0.1cm} 
  \begin{minipage}[b]{0.5\linewidth}
    \centering
    \includegraphics[width=6.5cm]{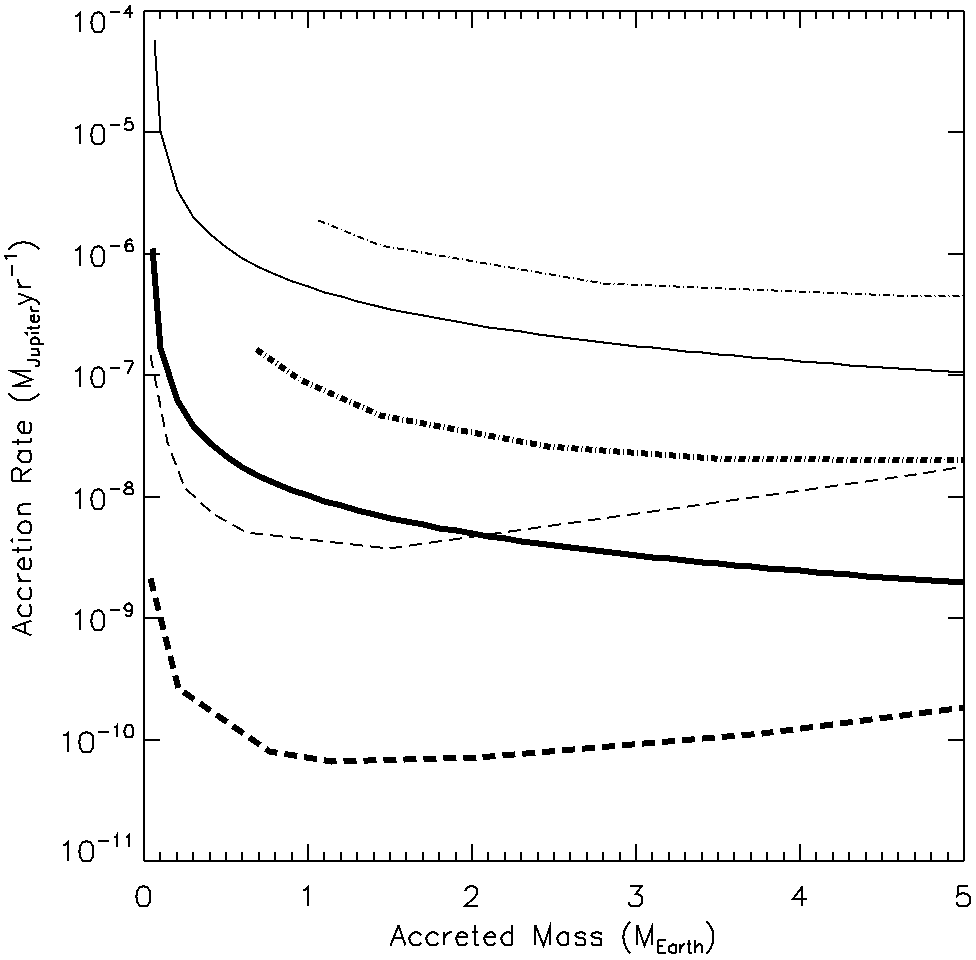}
    \caption{\textbf{a.}~Protoplanet accretion rates. The asterisks mark the accretion rates for our SPH calculations using a locally-isothermal equation of state, essentially providing upper limits to the accretion rates obtainable with various protoplanet masses. The diamonds mark the results of \protect \cite{batezeus}, and the plus signs those of \protect \cite{LubSeiArt1999}, connected with solid lines. These calculations were also locally-isothermal but were global disc simulations performed using the ZEUS code.  The SPH and ZEUS accretion rates are in reasonable agreement.  The accretion rates from our self-gravitating radiation hydrodynamical SPH calculations using core radii of 1\% of the protoplanet's Hill radius are given using line types that denote the different grain opacities.  Results are shown using standard interstellar grain opacities (IGO) (dotted), 10\% IGO (short-dashed), 1\% IGO (dot-dash), and 0.1\% IGO (long dashed).  The inclusion of radiative transfer substantially lowers the accretion rates of low-mass protoplanets, but Jupiter-mass protoplanets have similar accretion rates to the locally-isothermal result, regardless of the grain opacity. The analytic approximation of \protect \cite{angelo2003} is shown by the solid curved line. \textbf{b.}~Accretion rates versus the gas mass accreted by planetary cores for 1\% (thin lines) and standard (thick lines) grain opacities.  We plot the results of \protect \cite{papa2005} for 5~M$_{Earth}$ (dashed lines), and 15~M$_{Earth}$ (dot-dashed lines) cores. The solid lines show the extrapolations of our 10~M$_{Earth}$ core results.  Our 10~M$_{Earth}$ core accretion rates lie between the accretion rates obtained by \citeauthor{papa2005} for their 5~M$_{Earth}$ and 15~M$_{Earth}$ cores.  The effect of the reduced opacities is also consistent with \citeauthor{papa2005}'s results.}
    \label{figure1}
  \end{minipage}
\end{figure}

\begin{figure}
  \includegraphics[height=.9\textheight]{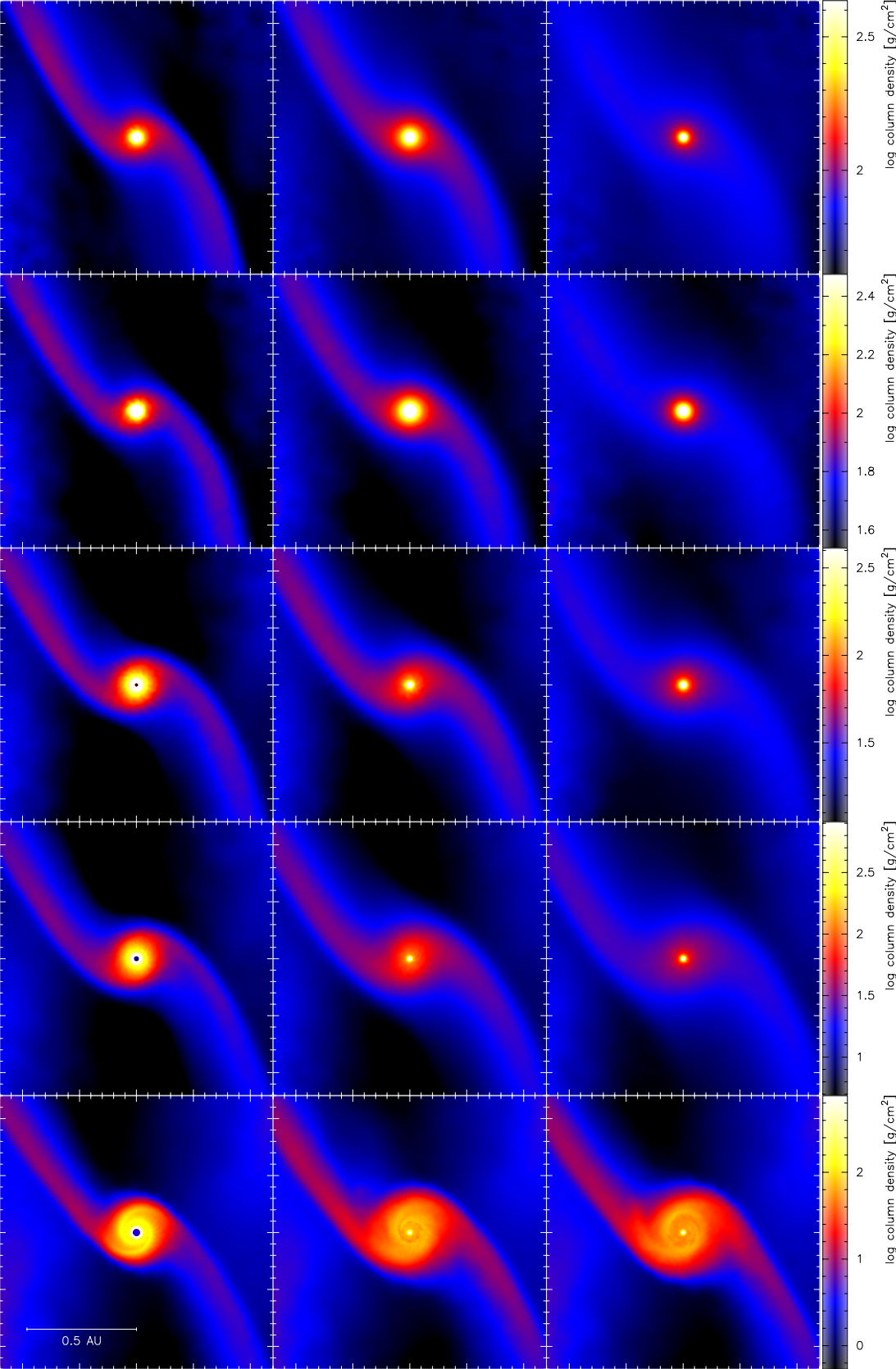}
\caption{Surface density plots for locally-isothermal calculations (left), and self-gravitating radiation hydrodynamical calculations using 1\% IGO (centre), and standard IGO (right) calculations with our standard protoplanetary disc surface density. From top to bottom the protoplanet masses are 22, 33, 100, 166 and 333 $\rm M_{Earth}$ respectively.  The radiation hydrodynamical calculations use protoplanet radii of 1\% of the Hill radii, while the locally-isothermal calculations use 5\% of the Hill radii.  Note that with radiative transfer and standard opacities, the spiral shocks in the protoplanetary disc are much weaker than using a locally-isothermal equation of state, while using radiative transfer with reduced grain opacities results in intermediate solutions because the discs are less optically thick and are able to radiate more effectively than with standard opacities.}
\label{figure2}
\end{figure}



\bibliographystyle{aipproc}   

\bibliography{ayliffe}

\IfFileExists{\jobname.bbl}{}
 {\typeout{}
  \typeout{******************************************}
  \typeout{** Please run "bibtex \jobname" to optain}
  \typeout{** the bibliography and then re-run LaTeX}
  \typeout{** twice to fix the references!}
  \typeout{******************************************}
  \typeout{}
 }

\end{document}